
\centerline{\bf PRECISE VACUUM STABILITY BOUND}
\centerline{\bf IN THE STANDARD MODEL}
\vskip 2 cm
\centerline{\bf addendum}
\vskip 1 cm
\centerline{Marc Sher}
\centerline{Physics Dept., Coll. of William and Mary}
\centerline{Williamsburg, VA  23187}
\vskip 1 cm
The results of this paper (Phys. Lett.  B317, 159 (1993)) were
given for top quark masses between $120$ and $160$ GeV.  Here, I
include the results for masses between $160$ and $190$ GeV.  The
procedure is unchanged  The results, corresponding
to Eq. 8, are now given by
$$m_H>132+2.2(m_t-170)-4.5\left({\alpha_s-0.117\over 0.007}\right)
$$
in units of GeV.  This matches the previous result for $m_t=160$
GeV.  The slightly greater dependence on the strong coupling is
due to the larger Yukawa coupling.  This is accurate to $1$ GeV
in the top mass and $2$ GeV in the Higgs mass.  Thus, discovery
of the Higgs boson in the next few years at LEP 200 would rule
out the standard model if one requires that the vacuum be stable
up to very large scales (we used the scale of $10^{15}$ GeV--the
precise value doesn't much matter).  If the vacuum is unstable,
the early universe would enter the true vacuum (ruling out the
model) unless the Higgs mass is within $5$ GeV of the bound, as
indicated in Fig. 2 of the paper.
\end